\documentclass[acmconf,authorversion]{acmart}

\AtBeginDocument{%
  \providecommand\BibTeX{{%
    \normalfont B\kern-0.5em{\scshape i\kern-0.25em b}\kern-0.8em\TeX}}}

\copyrightyear{2021}
\acmYear{2021}

\acmBooktitle{}
\acmPrice{}
\acmISBN{}

\usepackage{color}
\usepackage{xcolor}
\definecolor{myblue}{RGB}{179,205,227}
\definecolor{myred}{RGB}{251,180,174}
\definecolor{mygreen}{RGB}{204,235,197}
\definecolor{textred}{RGB}{228,26,28}
\definecolor{textblue}{RGB}{55,126,184}
\definecolor{textgreen}{RGB}{77,175,74}
\usepackage{url}
\usepackage{enumitem}
\usepackage{multirow}
\usepackage{wrapfig}

\begin{document}


\title{Visualizing Rule Sets: Exploration and Validation of a Design Space}



\author{Jun Yuan}
\affiliation{%
  \institution{New York University}
  \city{New York}
  \country{USA}}
\email{junyuan@nyu.edu}

\author{Oded Nov}
\affiliation{%
  \institution{New York University}
  \city{New York}
  \country{USA}}
\email{onov@nyu.edu}

\author{Enrico Bertini}
\affiliation{%
  \institution{New York University}
  \city{New York}
  \country{USA}}
\email{enrico.bertini@nyu.edu}

\renewcommand{\shortauthors}{Yuan, et al.}

\begin{abstract}
  Rule sets are often used in Machine Learning (ML) as a way to communicate the model logic in settings where transparency and intelligibility are necessary. Rule sets are typically presented as a text-based list of logical statements (rules). Surprisingly, to date there has been limited work on exploring visual alternatives for presenting rules. In this paper, we explore the idea of designing alternative representations of rules, focusing on a number of visual factors we believe have a positive impact on rule readability and understanding. The paper presents an initial design space for visualizing rule sets and a user study exploring their impact. The results show that some design factors have a strong impact on how efficiently readers can process the rules while having minimal impact on accuracy. This work can help practitioners employ more effective solutions when using rules as a communication strategy to understand ML models.
\end{abstract}

\ccsdesc[500]{Human-centered computing~Empirical studies in visualization}
\ccsdesc[500]{Human-centered computing~Information visualization}

\maketitle
\setlist[enumerate]{label*=\arabic*.}

\section{Introduction}

\texttt{If-then} rules are common in our daily lives. People use \texttt{if-then} rules to make decisions, such as "\textit{If it is sunny today, then I will go to the grocery.}" People use \texttt{if-then} rules to make predictions, such as "\textit{If we see red sky in the morning, then it will rain today.}" And with the high demand for explanations of machine learning models, rules in the \texttt{if-then} format are also widely used  as a way to understand the logic of how a model works.


Rule-based machine learning (ML) models have been used for decades in decision-making and prediction~\cite{clark1991rule,liu1998integrating,evans2018learning,lakkaraju2016interpretable,letham2015interpretable,safavian1991survey,wang2017bayesian,yang2017scalable}. Although complex models such as deep neural networks (DNN) and ensemble models usually have higher performance, they are also limited by their complexity as well as low transparency and intelligibility; making their use limited in settings where transparency and user trust are important (e.g., healthcare, security, public policy). In recent years, the increasing need for model interpretability and explanation has led to a resurgence of rule-based models; with algorithms that aim at achieving similar levels of predictive power as less transparent solutions~\cite{lakkaraju2016interpretable, wang2015falling,wang2017bayesian}. Rules are also being used as a way to describe the behavior of existing black-box models by using model-agnostic explanation~\cite{ribeiro2018anchors,guidotti2018local,sanchez2015towards,lakkaraju2019faithful}. Once rules are generated, they are generally shown as a list of logical statement expressed with text. These lists can easily become long and complicated and, as such, hard to read and understand. An open question in this area is whether adopting alternative visual representations may lead to improvements in terms of rule sets readability and understanding.

While initial attempts to produce more effective visual representations of rules have been made~\cite{ming2019rulematrix,soares2020explaining,di2019surrogate}, progress in this area has been surprisingly limited in scope and not sufficiently systematic. More specifically, while prior work proposes alternative ways to visualize rules, we are not aware of systematic analyses of how rules \textit{could} be visualized and what factors may play a role in their effectiveness. This is important because a systematic analysis of the visual representation space can help researchers and designers think more productively about potential solutions and make more informed decisions. Currently, those who want to adopt rules, either as a way to explain existing models or as a way to build models that are inherently interpretable, lack guidance on how to communicate the results visually and effectively. \looseness=-1






In this work, we make progress in this direction by proposing an initial set of \textit{visual factors}: ways in which a given set of rules can be presented visually according to how the elements of a rule can be arranged spatially and how their properties can be encoded visually. We then focus on two specific factors: \textit{feature alignment} (weather the features of a rule are organized in an aligned or unaligned layout)  and \textit{predicate encoding} (whether the predicates are presented through symbolic or graphical depictions) and study them in detail through a crowdsourced controlled experiment.






With relevant details presented in the rest of the paper, our study confirms a strong impact of the two factors. Both alignment and graphical encoding decrease response time in rule reading tasks, with no negative impact on accuracy. Alignment with a very strong effect size and graphical encoding with a much more moderate effect.



The contribution of this paper is threefold. First, we propose a characterization of visual factors for visual encoding of rule sets. This characterization can help in the design of novel representations for rules and in the study of their impact on understanding. Second, we propose a hierarchical task analysis (HTA) that simulates the way readers process sets of rules visually. As we will demonstrate, the HTA is a very powerful tool to single out low-level perceptual tasks for which visual design interventions can provide improvements. Third, we contribute a controlled experiment that provides initial evidence for the positive impact of the solutions we propose and as such it provides empirically validated guidelines for rule visualization.







\section{Rule Sets: Background and Challenges}
\label{sec:background}
\begin{wrapfigure}{r}{.56\textwidth}
    \centering
    \includegraphics[width=0.52\textwidth]{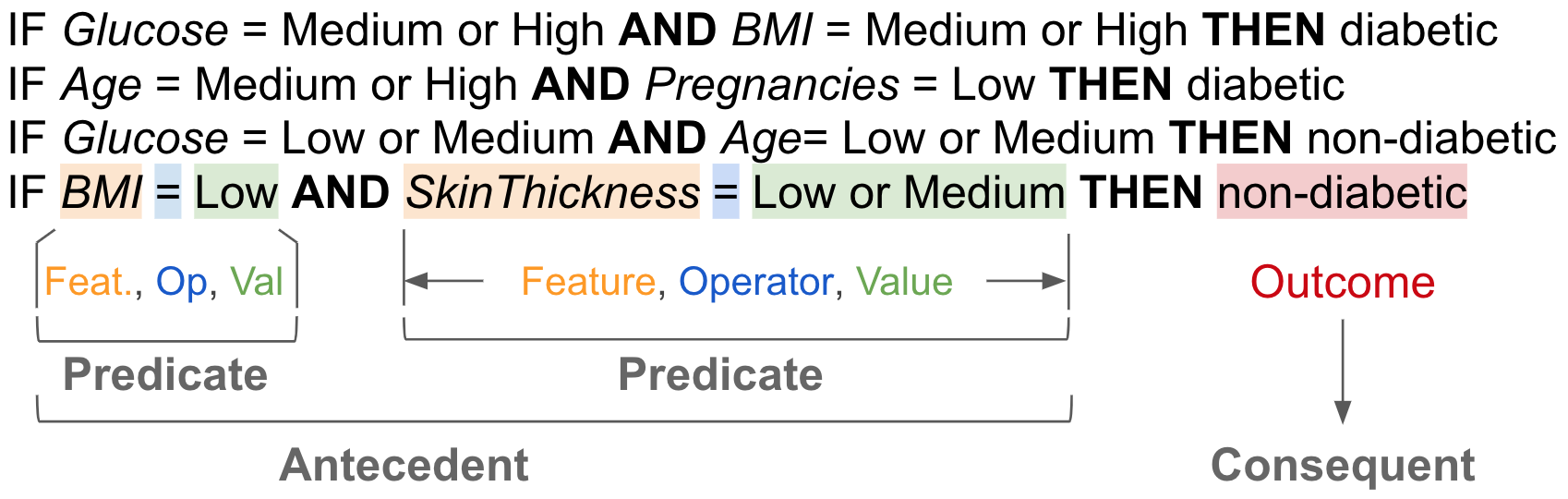}
    \caption{Terminology used to describe rules. Using rule set generated on the diabetes data set as an example. The antecedent can have multiple predicates. The consequent has one predicate which we call outcome.}
    \label{fig:pima}
\end{wrapfigure}

Rules have been used extensively in Machine Learning (ML) as an ``interface'' to communicate information about the logic used by a model (either for descriptive or predictive purposes). Many different types of rules exist in ML which capture different types of patterns. In this work we focus exclusively on \textit{classification rules}; rules that describe associations between the feature values found in the data and an outcome of interest (e.g., health indicators and risk of becoming diabetic). Rules are structured as logical statements such as those presented in Figure~\ref{fig:pima}. A rule is structured as a combination of predicates representing the \textit{antecedent} of the rule and a single predicate representing the \textit{consequent} of the rule, which we call \textit{outcome}. A rule is meant to describe a logical inference for which if the antecedent of a rule is true then the consequent is true.






Rules can be generated in many different ways and can be organized in different types of structures. Rule lists~\cite{clark1991rule} are organized in a ``cascading structure'' in which the order of rules matters, that is, a rule is ``active'' only when the conditions included in the preceding rules are not met~\cite{lakkaraju2016interpretable}. Decision trees organize rules in a tree structure in which each node represents one predicate and a path from the root to a leaf represents a whole rule. Finally, rule sets produce sets of independent rules in which the order of rules is not meaningful. In the following we will focus exclusively on rule sets, leaving the study of other structures to future work.

\subsection{Use of Rules in Machine Learning}

The concomitant increase in complexity and widespread adoption of machine learning models has led to a more pressing need and interest in machine learning interpretability. Models need to be intelligible for developers and users; especially in domains where sensitive decisions are made. Rules are used in two main contexts: building rule-based models and building explanation models to describe the behavior/logic of trained black-box models. Predictive rule-based models have been used for a long time in ML, but their use has been hindered by their poor performance compared to other more sophisticated and less transparent solutions~\cite{safavian1991survey,clark1991rule, liu1998integrating}. With a stronger need for interpretability, researchers recently started developing new rule algorithms that produce better performance than previous methods~\cite{evans2018learning,lakkaraju2016interpretable,letham2015interpretable,wang2017bayesian,yang2017scalable}, and which employ strategies to make them more interpretable~\cite{lakkaraju2016interpretable}. Many explanations models have been introduced in recent years~\cite{guidotti2018survey} and in this context rules play a major role. Craven and Shavlik, in their seminal work on surrogate models, developed an algorithm to create intelligible descriptions of neural networks~\cite{craven1996extracting}. More recently, researchers have developed innovative methods to create rules that describe a model globally~\cite{sanchez2015towards} or even locally, for a small set of instances~\cite{ribeiro2018anchors, guidotti2018local}. Whether rules are used as a predictive model or as a description (explanation) of an existing one, they typically have a very similar structure, therefore the work we propose in this paper applies to both cases.





\subsection{Rule Complexity and Understanding}
\label{sec:rule-complexity}

Despite the fact that rules are widely considered ``interpretable'' due to their intuitive and transparent representation~\cite{molnar2019}, their use does not necessarily guarantee that a model can be easily understood. As the number of rules and/or the number of predicates increases, reasoning about the model's logic becomes extremely hard. Even basic operations like finding rules that share a common set of conditions can become highly effortful and error prone with more than a handful of rules (lists with more than about 10 rules can become surprisingly daunting to interpret).

Rule intelligibility can be expected to degrade as rule complexity increases. Recently researchers identified a number of factors that influence complexity with different names~\cite{lage2019evaluation,lakkaraju2016interpretable, wang2017bayesian}. Some of the most common include: the total number of attributes used in the predicates contained in the rule list (dimensionality), the total number of rules (cardinality), the amount of overlap between rules (redundancy), the maximum number of predicates that a rule can include (maximum rule length), the number of distinct values each predicate can have (feature cardinality). While a complete investigation of how these factors impact rule understanding does not exist yet, some few empirical studies on rule structure exist.





    


Huysmans et al., evaluated the performance of several cognition-related tasks with rules of size over 7 levels~\cite{huysmans2011empirical} where people tend to spend longer time with more complex tasks. More recently, Lage et al., studied how rule complexity influences rule interpretability~\cite{lage2019evaluation} and found that different types of complexity such as repeated terms and rule size can influence the response time, accuracy and confidence in 3 tasks across different domains. In a recent paper, Lakkaraju et al., compared Bayesian rule lists~\cite{letham2015interpretable} with \textit{decision rule sets}~\cite{lakkaraju2016interpretable} and found that the \texttt{if-then} structure of rule sets, as opposed to the \texttt{if-then-else} structure of rule lists, improves the accuracy and efficiency of rule interpretation. A few studies also exist on comparing hierarchical versus flat (list) organizations of rules. Subramanian et al., investigated how well people can simulate investment strategies (decision making) showing rules either in a tabular or hierarchical format and found that trees are a better for decision support than a hierarchical table for interpreting conditional logic~\cite{subramanian1992comparison}. Another similar study compared decision trees, structured \texttt{if-then-else} rule lists, and tables~\cite{vessey1986structured} and found that structured text outperformed decision trees and tables when the goal was to identify specific conditions to match an action. Finally, in the same work we cited above~\cite{huysmans2011empirical}, three different representations of rules have been tested: decision tables, decision trees, and lists. The results showed that table led to the best performance. While most of these studies focus more on the complexity of the information produced by rule algorithms and their impact on understanding, our focus is on how visual representation may impact understanding while keeping this information a fixed factor. 


\subsection{Research on Rule Sets Visualization}

In the studies reviewed above, the only ones that study aspects related to visual representation are those that compare hierarchical versus flat representations of rules. An exclusive focus on rule structure is, however, limited because it is not always possible to structure the output of an existing rule algorithm into a hierarchy. Exploring different visual encoding strategies can potentially enhance readability and improve understanding. While visual designs that follow this intuition have been developed recently, this nascent area of research needs much more work to understand what the opportunities in this space are. \textit{RuleMatrix}~\cite{ming2019rulematrix} uses a matrix layout and graphical encoding of predicates. \textit{Explainable Matrix}~\cite{neto2020explainable} uses a similar metaphor but extends it to increase its scalability. \textit{MegaClouds}~\cite{soares2020explaining}, uses line symbols and icons to visualize the range of values represented by a rule predicate. To the best of our knowledge, none of these solutions have been tested in controlled experiments. In addition we are not aware of a systematic exploration of the design space of solutions for rule visualization. These two gaps are the main motivation for this work in which we aim at exploring rule visualization more systematically and at producing initial empirical evidence for the effect of selected visual factors. \looseness=-1


\section{Visual Factors for Representation of Rule Sets}
\label{section:factors}

We now more systematically explore visual factors that can be used to design alternative visualizations of rule sets. In doing that, it is important to clarify the scope of our work in suggesting these factors. We focus exclusively on flat representations because not all rule algorithms produce hierarchical structures. Conversely, hierarchical rules can always be transformed into a list of rules. Moreover, we focus exclusively on \textit{visual} representation of rule content, not on how rule properties may impact understanding (e.g., rule dimensionality and cardinality as described in Section~\ref{sec:background}). 
Finally, we explicitly focus on the problem of visualizing the \textit{logic} of rules, but not their statistical properties. So we aim at devising visual representations that enable complex logical inferences across rules. For example, we do not aim at visualizing rules in terms of statistical properties such as \textit{support} and \textit{confidence}; rather we aim at visualizing rules so that one can generate inferences about existing relationships between data features and outcomes. We will provide more details about this type of tasks in the following section, which is entirely devoted to our task analysis.
\begin{figure*}
    \centering
    \includegraphics[width=\textwidth]{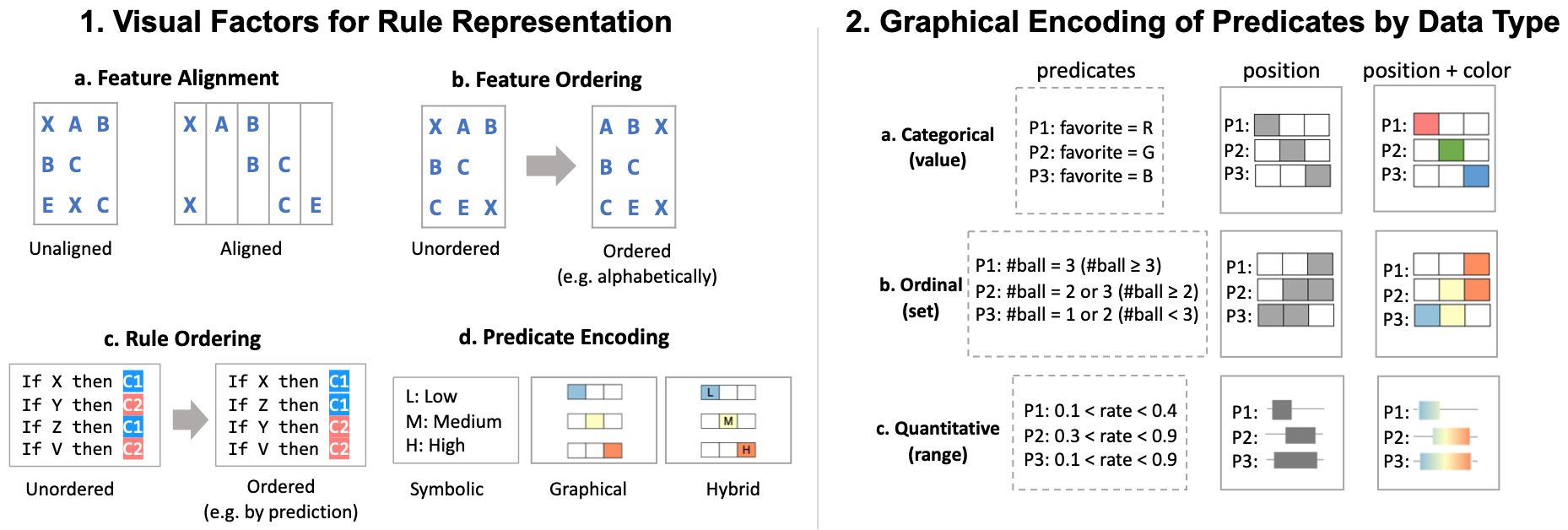}
    \vspace*{-.6cm}
    \caption{Part (1) includes four visual factors we identified that play a role in the process of human understanding a set of rules: a. Feature Alignment, b. Feature Ordering, and c. Rule Ordering are relevant to the \textit{Spatial Arrangement} of predicates in the presentation; d. \textit{Predicate Encoding} include the strategies of how to express the predicates in rules. Part (2) include examples of graphical encoding strategy for predicate by data type.}
    \vspace*{-.3cm}
    \label{fig:visual_factor}
\end{figure*}


Visual factors for rules can be grouped into two main classes: \textbf{Spatial Arrangement} and \textbf{Predicate Encoding}. Spatial arrangement refers to the problem of how to position the rule components in the visual space (layout) and visual encoding is about how to create visual representations of the values and conditions expressed in the rule's predicates. Figure~\ref{fig:visual_factor} provides an overview of the factors we have identified, which we describe in more details below. We also created a website\footnote{\url{https://rule-logic-vis.herokuapp.com/}} to present the rule visualization of all the visual factors we identified. \looseness=-1


\textbf{Feature Alignment.} In a standard textual representation the features referenced in each predicate are \textit{unaligned}, that is, they have a different location. See for instance Figure~\ref{fig:visual_factor}-1a:Unaligned, where predicates using the same features (X, A, B) may be located in different horizontal positions. An alternative arrangement is to have the predicates \textit{aligned} into a tabular format as shown in Figure~\ref{fig:visual_factor}-1a:Aligned. The unaligned arrangement is more compact but the aligned arrangement affords much easier comparison across the rules.

    
    
\textbf{Feature Ordering.} Features and their corresponding predicates can also be ordered according to different criteria (e.g., alphabetically) (see Figure~\ref{fig:visual_factor}-1b). In unaligned designs rules can be ordered individually. In aligned design they can be ordered by column. Feature ordering is useful because it can help give prominence to certain features over others. For instance, the columns of a table can be ordered according to how often a feature is used in the rules; thus making the most important features more prominent.
     

\textbf{Rule Ordering.} Rules can also be ordered according to relevant criteria. In the mock-up we show in Figure~\ref{fig:visual_factor}-1c, the rules can be ordered based on the consequent/outcome part of the rule. Rule ordering can make relationships between features and outcome much easier to detect. For example, if the rules are ordered by outcome, it is much easier to detect correlations between features and outcome.

    


\textbf{Predicate/Outcome Encoding.} The predicates of a rule are logical statements over a set of features, each with an associated domain of values which can be either \textit{categorical}, \textit{ordinal} or \textit{quantitative} (Figure~\ref{fig:visual_factor} Part 2). 
They can present a single value or multiple values or value ranges connected by logic $AND$ or $OR$ connectives (see the categorical, ordinal and quantitative example in Figure~\ref{fig:visual_factor}-2a,2b,2c)). The outcome is itself a predicate, typically with a single feature of categorical or ordinal type. \looseness=-1




When confronted with the problem of visually representing predicates we have identified three main broad strategies which are exemplified in Figure~\ref{fig:visual_factor}-1d. \textbf{Symbolic encoding} is the traditional way of representing rules through a mix of logical and textual symbols. \textbf{Graphical encoding} is the idea of representing predicates through graphical representations. \textbf{Hybrid encoding} is the idea of integrating both symbolic and graphical encoding in one solution.

The problem of how to create graphical representations of rule predicates is an interesting one because predicates are not just data values within a given domain but they represent ``constraints'' over a domain. In Figure~\ref{fig:visual_factor}-2 we propose a number of potential encoding strategies one can use to visualize predicates. The strategies are organized by data type (in rows) and visual channels (in columns). The first column uses simple text to explain the meaning of the predicates and the following columns propose alternative graphical strategies. To design these representations we took inspiration from the ranking of visual variables~\cite{bertin2011graphics}. Position (spatial location) is widely considered the most effective visual property to represent information visually. For this reason in our designs position (represented by the position of the small filled cells) is used to convey information about which elements in a set (or range) match the condition expressed by the predicate. In doing so, position expresses information about sets, ranges and sequences, which would otherwise be hard or even impossible to express exclusively with other visual channels (e.g., shape, orientation, size). In the second design we integrate color as redundant encoding with the information expressed by position. This solution is included because color, when used in conjunction with position, makes it easier to visually associate similar or identical values across multiple rules. With color it is possible to quickly identify rules that are \textit{qualitatively} on a different range of values, e.g., rules with very high or very low values for a given feature.

When predicates are visualized through a symbolic representation, as it is normally done in text-based rule sets, values across rules are evaluated exclusively through the meaning of their textual symbols. The advantage of using symbolic encoding is that they have a direct way of conveying meaning. For example, in Figure~\ref{fig:visual_factor}-1d:Symbolic, a low value is represented as an 'L', medium as an 'M', and high as a 'H'. If symbols are chosen in ways that are evocative of their meaning (or are just full words), then the association between symbols and meaning is effortless. However, symbolic encoding is also limited in terms of visual search and comparison, especially when the information is quantitative. For example, finding all rules with a given value range for \textit{Age} is slow and effortful. Graphical encoding conversely can speed up visual search and comparison~\cite{paivio1975perceptual} and thus make rule reading way less effortful.

The open issue we are faced with is therefore how to balance the benefits of symbolic encoding and those of graphical encoding. In theory, symbolic encoding should be slower and more effortful than graphical encoding, especially as the number of rules increases. However, the unfamiliar nature of graphical encoding may slow down viewers due to increased uncertainty, less precision and lack of familiarity with the medium. The main goal of the study we will present in Section~\ref{section:experiment} is to start gathering evidence about how these strategies compare when used to perform complex tasks with lists of rules.

\section{Task Analysis}
\label{sec:task-analysis}

If we want to understand the impact visual factors have on performance, it is necessary to have a specific description of what visual operations users carry out when reading rule sets. In this section we propose a hierarchical task analysis (HTA) of the two main tasks we believe represent the main operations performed with rule sets. We then use the HTA to build hypotheses about how the factors we have identified may impact performance. These hypotheses will be the basis for the study we present in the following section.

Our analysis starts from the conjecture that the main goal of rules is to perform two main types of inferences: \textit{prediction estimation} and \textit{prediction characterization}, respectively drawing inferences from the \textit{antecedent} to the \textit{consequent} part of the rules for estimation and the other way around for characterization.

\textbf{Prediction estimation.} In prediction estimation the main goal is to estimate what is the most likely outcome for an instance with a given set of values. For example, using the diabetes prediction scenario mentioned before one question of this type could be: ``\textit{What is the most likely prediction for a person who has a \textit{BMI} of 50 and \textit{Age} equal to 48?}''. It's important to keep in mind that people typically have only partial specifications for the values of instances they are interested in (imagine a physician thinking of prototypical patients with certain characteristics). This means that partial specifications, in the more general case, may match only a subset of the predicates of the rules and, as a consequence, they may match multiple rules at once. When this happens, the type of inference necessary to estimate the most probable outcome involves comparing outcomes from individual rules and producing a synthesis from their comparison (we provide more details below in the HTA).

\textbf{Prediction characterization.} In prediction characterization the main goal is to build explanations of what are prototypical instances for a given outcome. For example, one question of this type could be: ``\textit{What kind of characteristics should a patient have to be predicted as at high risk of diabetic?}''. That is, starting from rule outcomes, the goal is to build prototypes that link data values to a target outcome. This type of inference also requires comparison between rules. Starting from the target outcome of interest, the reader needs to identify common patterns in the antecedent part of the rules with the outcome and develop a \textit{synthesis} of what are prototypical instances (e.g., patients) with that type of outcome. \looseness=-1

\begin{figure*}
    \centering
    \includegraphics[width=.9\textwidth]{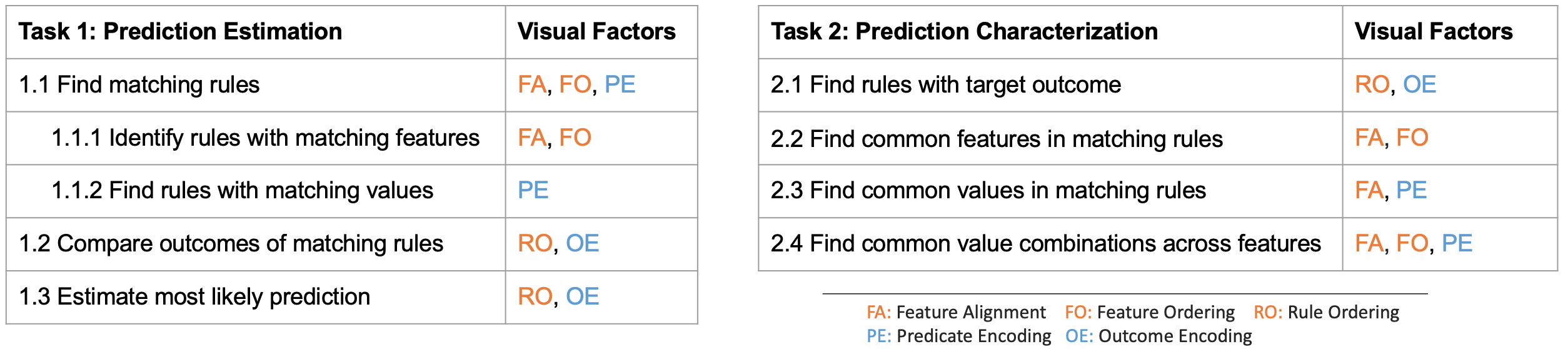}
    \vspace*{-.3cm}
    \caption{Hierarchical task analysis of rule understanding tasks. Visual factors that impact performance.}
    \label{fig:hta}
    \vspace*{-10pt}
\end{figure*}

In the following, we provide a more systematic analysis of these two main tasks. For each one we start with the high level goal and break it down into lower-level tasks to analyze what kind of visual operations are needed to perform the high-level task. We perform the HTA to identify more precisely which visual operations may benefit from the visual features we propose to analyze. The description of the two type of inferences we provided above is in fact too abstract to hypothesize when and why a visual representation could be reasonably expected to provide an advantage over another. Figure~\ref{fig:hta} provides a summary of the HTA as well as of which visual factors may impact performance in their execution.

To estimate prediction based on the available information (Task 1.1), a viewer needs to read across the rules and find the qualified rules that contain conditions that match the given feature values. To find rules that match the given information, a person needs to first identify which rules contain the \textit{features} that match the conditions (Task 1.1.1) and then find the predicates that match the \textit{value(s)} of the conditions (Task 1.1.2). Once the matching set of rules has been identified, the next step requires to compare the outcomes of the matching rules (Task 1.2) and finally estimate the most likely prediction (Task 1.3). This deconstruction into lower-level tasks enables us to reason about how the visual factors we described in Section~\ref{section:factors} may impact performance in performing this task. As summarized in Figure~\ref{fig:hta}, each sub-task can be influenced by one or more visual factors.

\begin{figure*}
    \centering
    \includegraphics[width=\textwidth]{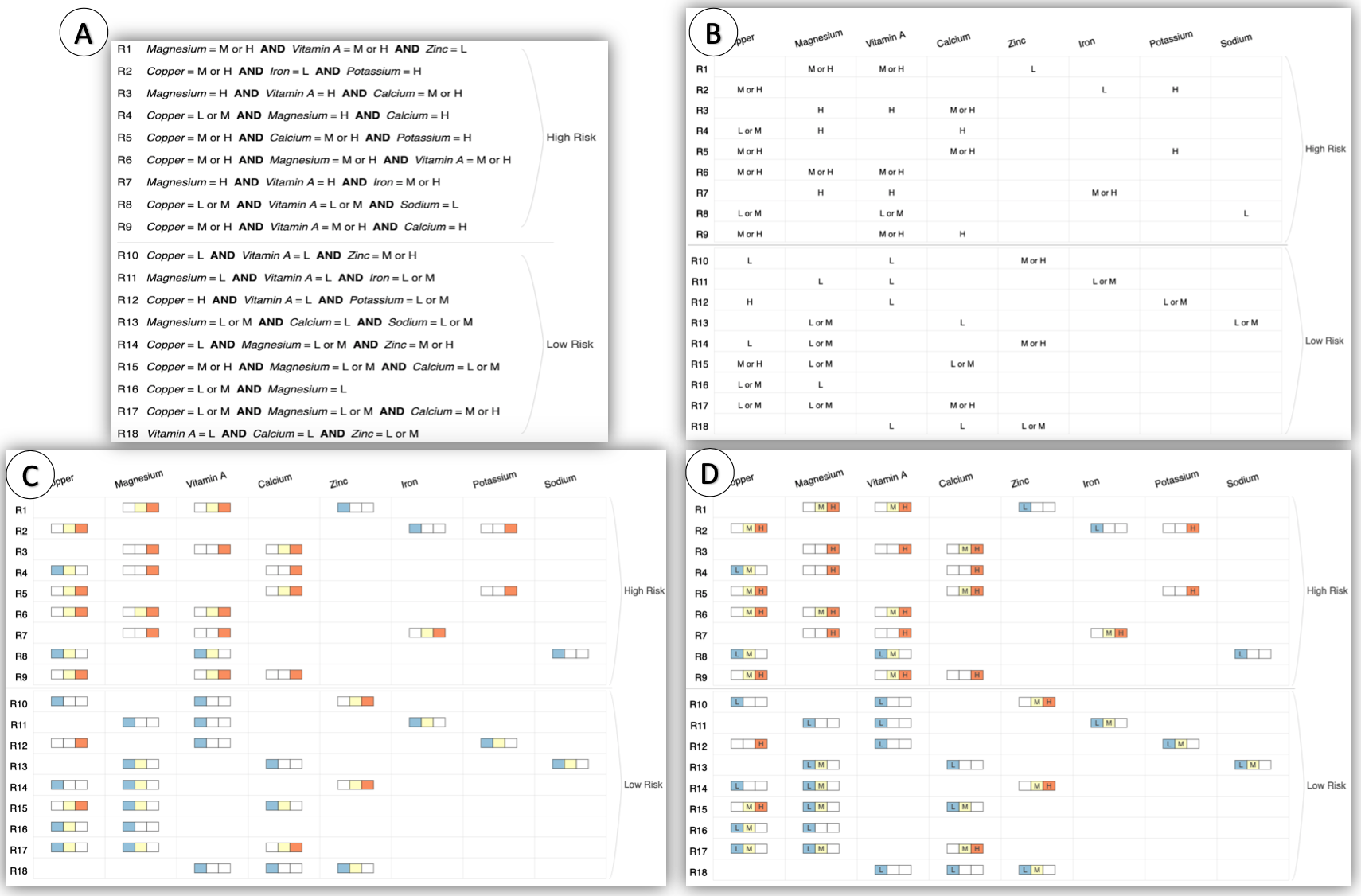}
    \caption{The four visual representations that are tested in the study. (A) feature unaligned, symbolic encoding; (B) feature aligned,  symbolic encoding; (C) feature aligned, graphical encoding; (D) feature aligned, hybrid encoding.}
    \label{fig:test_vis}
    \vspace*{-5pt}
\end{figure*}

Feature \textit{alignment (FA)} and \textit{feature ordering (FO)} can impact Task 1.1.1 by making it easier to verify which rules contain a given set of features. If the features are not aligned, then every time a viewer visually scans a new rule they have to verify whether the set of features of interest is present or not; whereas with an aligned layout such operation is straightforward because all features are aligned in the same position. Alignment has its own drawbacks also: when features are aligned, the overall layout of the rule presentation is less compact and reading individual rules may be harder. For some rules the predicates may be far from each other making it harder to identify which predicates belong to a given rule. An additional advantage of alignment and ordering is that features can be ordered according to some useful criteria. For instance features can be ordered according to how frequently they are used in the rule set, thus making it easier to concentrate the attention on smaller area of the visualization rather dispersed across many columns. To get a sense of how alignment and sorting may impact this task you can compare the two visualizations A and B in Figure~\ref{fig:test_vis}. \looseness=-1




Task 1.1.2 is more impacted by \textit{predicate encoding (PE)}, that is, which strategy is used to visualize the predicate condition. When a viewer needs to verify whether a predicate satisfies a given condition (e.g., whether the predicate includes a given value or not), the way the predicate is represented may impact how easy it is to perform such verification. In Figure~\ref{fig:visual_factor}-1d, we presented some potential options showing how the same predicate may be visualized with different encoding strategies. If the predicate is represented with text, the viewer will need to look for the term or symbol that matches the condition. If the predicate is represented using a graphical depiction, the same operation will be performed through graphical reasoning. Finally, if the predicate is represented using a hybrid encoding we can expect a mix of the two strategies to take place. Which one among these works best is an open question and one of the main research questions we want to address. \looseness=-1



Finally, for Task 1.2 and Task 1.3 (comparing outcomes and estimating the prediction) we expect \textit{rule ordering (RO)} and \textit{outcome encoding (OE)} to have an impact on performance. Rules can be ordered according to many different criteria and each ordering may hinder or facilitate the comparison of the outcomes of the matching rules. For instance, ordering the rules by outcome values makes it easier to quickly identify how many matching rules belong to a given outcome or to another (e.g., see the ordering of rules in Figure~\ref{fig:visual_factor}c). Outcome encoding may also make the comparison and synthesis of outcome values more or less effective. Since outcome is itself a (single) predicate, the same reasoning we applied to visual encoding of predicates above applies to outcome encoding.




To understand how common characteristics of feature values may lead to a given outcome in \textit{prediction characterization}, we expect the viewer to go through the following sub-tasks. First, one needs to single out rules that contain the outcome of interest (Task 2.1). Once these rules have been detected the next step consists of identifying common features (Task 2.3) and common predicate values (Task 2.4). Finally once these common values have been identified for individual features, the viewer will identify combinations of common values and features across multiple features (Task 2.5). Similarly to what we have described for Task 1, here we describe how our visual factors impact performance of these sub-tasks (see summary in Figure~\ref{fig:hta}).

For Task 2.1, we expect \textit{rule ordering (RO)} and \textit{outcome encoding (OE)} to influence performance. For Task 2.2-2.4 we expect \textit{feature alignment (FA)} to have an impact since they involve comparison across multiple rules. We also expect \textit{predicate encoding (PE)} to play a role in Task 2.3. and Task 2.4 since they involve the comparison of values within the predicates. Finally, for the same reasons outlined in Task 1, we expect \textit{feature ordering (FO)} to have an impact on Task 2.2 and Task 2.4. \looseness=-1

\section{Experiment}
\label{section:experiment}
To better understand the impact of the identified visual factors we designed a controlled experiment that aims at teasing out their effect on rule understanding. In this section we describe the design of the experiment and report on the results.


%

\subsection{Experiment Design}


\subsubsection{Stimuli: Visual Representations and Rule Sets.}
The study is organized as a controlled experiment in which we compare conditions that vary according to the factors we want to study, namely: the alignment strategy (aligned and unaligned) and the predicate encoding strategy (symbolic, graphical, and hybrid), as described in Section~\ref{section:factors}.  

The main goal of the study is to understand the impact our interventions have with respect to the traditional representation of rules based on text (see Figure ~\ref{fig:test_vis}A). For this reason our study uses a standard textual list of rules as a \textit{control condition} and organizes the study around improvements over this control condition. In the study, we include the following four conditions:

\begin{itemize}[noitemsep,topsep=1pt]
    \item[(\texttt{SU})] \textbf{S}ymbolic encoding for predicates, \textbf{U}naligned features (baseline, control condition);
    \item[(\texttt{SA})] \textbf{S}ymbolic encoding for predicates, \textbf{A}ligned features;
    \item[(\texttt{GA})] \textbf{G}raphical encoding for predicates, \textbf{A}ligned features;
    \item[(\texttt{HA})] \textbf{H}ybrid encoding (graphical and symbolic) for predicates, \textbf{A}ligned features.
\end{itemize}

The \texttt{SA} condition is used to single out the effect of alignment. In fact \texttt{SU} and \texttt{SA} differ only in the alignment strategy they use, leaving all the rest equal. \texttt{GA} and \texttt{HA} are used to single out the effect of graphical and symbolic encoding. In fact, \texttt{GA} and \texttt{HA} differ only in terms of predicate encoding strategy employed with respect to \texttt{SA}. The actual conditions we tested in the study are shown in the Figure~\ref{fig:test_vis}. Having to choose a specific data type to cover, we opted for ordinal data since it covers useful properties of both categorical and quantitative encodings (see Figure~\ref{fig:visual_factor}-2). Since studying sorting strategies for the rule is not an objective of the study, we also fixed the order of the rules and decided to sort by outcome as shown in the figures (\textit{High Risk} on top and \textit{Low Risk} at the bottom). Similarly, we also decided to fix the order of the features and opted to sort by frequency of features in the rule set; with frequency decreasing from left to right. \looseness=-1

For the purpose of this study we use two main data sets: one is a credit risk data set~\cite{fico} for the training phase of the study, to help the participants to familiarize with the tasks and concepts of the study, and another is the diabetes data set~\cite{smith1988using} for the actual study. \looseness=-1


To generate the rules, we used the algorithm proposed by Wang \textit{et al.}~\cite{wang2017bayesian}. For training, we needed a rule set that was simple enough to explain the visualization and complex enough to demonstrate the tasks. As for testing, we created a rule set with an higher level of complexity in order to differentiate the effects of the tested visual factors. The final set includes $18$ rules over $8$ features where each rule has a maximum of $3$ predicates in the antecedent. The consequent is a binary variable that can assume two values: high or low risk.

\looseness=-1


In designing the rules we also had to mitigate the effect of \textit{prior knowledge}. In one of our pilot studies, we tested the visualizations of rules generated from a diabetes data set~\cite{smith1988using} and found that the participants could predict the answer by using intuition and with little engagement with the visual representations. To mitigate this effect, we changed the feature names (originally \textit{Age}, \textit{BMI}, etc.) into names of minerals such as \textit{Magnesium}, \textit{Calcium}, \textit{Zinc}, etc., and invented an hypothetical disease. The subjects were instructed to analyze the predictions of a model that used the mineral features to predict the level of risk of contracting the hypothetical disease. The risk prediction was either \textit{low} or \textit{high}.


\subsubsection{Tasks.} Starting from the task analysis in the previous section we designed two sets of test tasks: one set for \textbf{Prediction Estimation (T1)} and one for \textbf{Prediction Characterization (T2)}. For prediction estimation we asked the participants to answer questions of the type: \textit{``What is the most common prediction for rules containing conditions that match a person with a High value of \textit{Calcium}?''}. For prediction characterization we asked questions of the type: \textit{``Considering only the rules that predict high risk, what is the most common value for Magnesium?''}. For both type of questions we introduced variations that increased complexity progressively. In order to vary the level of complexity we used two strategies. First, we increased the number of features involved in the questions. For example a more complex version of the first example above would be a question that includes two features: \textit{``What is the most common prediction for rules containing conditions that match a person with a High value of Calcium \textbf{and} a low value of Magnesium?''}. Second, we varied the questions so that the target range of a predicate would include more than one value, that is, instead of asking to find rules for a high value of \textit{Magnesium}, we would ask for rules with a \textit{medium or high} value of Magnesium. Similarly we found in pilot studies that medium values are harder to identify than extreme values and we used this fact to modulate the complexity of the tasks. With these variations in complexity we created a total of $10$ test questions: $5$ for Task 1 and $5$ for Task 2. More details can be found in the Github web page we created for this study \footnote{\url{https://github.com/nyuvis/rule_empirical_study}}.

\begin{figure*}
    \centering
    \includegraphics[width=\textwidth]{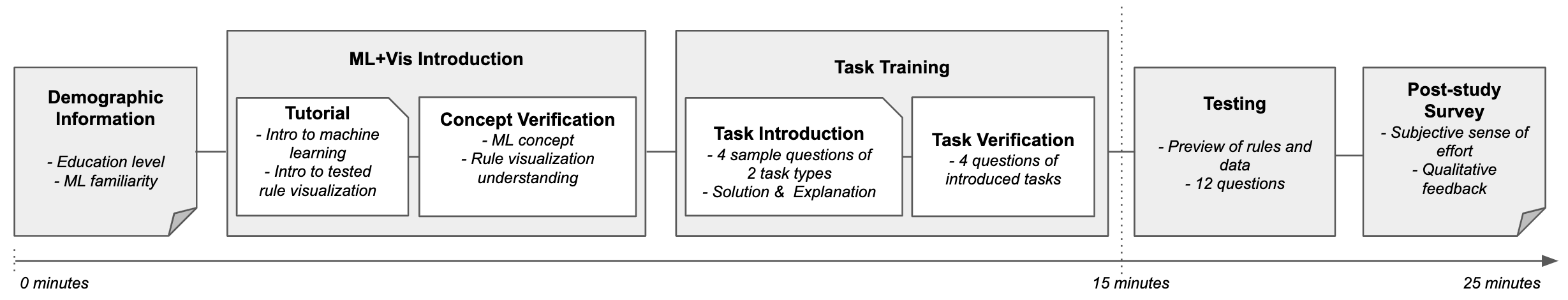}
    \vspace*{-.5cm}
    \caption{The study procedure consists of 7 stages: \textit{Demographic Data Collection}, \textit{Tutorial}, \textit{Concept Validation}, \textit{Task Introduction}, \textit{Task Validation}, \textit{Testing}, \textit{Survey}. }
    \vspace*{-.5cm}
    \label{fig:procedure}
\end{figure*}

\subsubsection{Procedure.}
We presented the 4 conditions outlined above to 4 separate groups of participants using a between-subjects design. The study was organized according to the following steps which were common to every group except the specific type of condition used in the study: (1) We started with a consent form and collection of \textit{demographic information}; (2) We asked the participants to watch a 4-minute \textit{tutorial} video to introduce basic ML concepts and to learn how to read the rule visualizations. This step was followed by a test to verify that the participant had learned the concept. This was used as a pre-requisite to move on to the next steps; (3) We described the tasks through a \textit{task introduction} page showing example tasks to let the participants familiarize with the test. Then, we asked them to perform a few simplified tasks resembling those used in the test, but using a different data set. We used this steps at an additional verification test to decide whether the participant could move on to the actual test; (4) In the \textit{test} stage, we first showed the visualization used in the test and described the test process. Then, for each task we asked the participants to answer 6 questions for each task type. Each question was posed as a choice in a multiple-choice test where only one out of three answers was right. For each question the participant also had to specify the confidence level associated to the answer. The values ranged between $1-5$ with $1$ being 'not confident at all' and $5$ being 'extremely confident'. The first question of each task type was used only for habituation to the task and was eliminated from the analysis. As a final step, after answering all the questions the participants had to answer a final question on their their overall subjective sense of effort in performing the tasks. 
The complete study procedure is shown in the Figure~\ref{fig:procedure}.
\looseness=-1





\subsubsection{Performance Metrics.}
To evaluate the performance of rule reading tasks, we used the following four metrics:
\begin{itemize}[noitemsep,topsep=1pt]
    \item \textit{Accuracy:} The number of correct answers over the number of questions.
    \item \textit{Response Time (RT):} The time required to answer a question.
    \item \textit{Confidence:} The subjective confidence scores collected for each question in a range $1-5$.
    \item \textit{Subjective Effort:} The subjective sense of effort scores collected for each question in a range $1-5$.
\end{itemize}

\subsubsection{Participants.}
We recruited our participants through the crowdsourcing platform Prolific~\footnote{\url{https://www.prolific.co/}}; a validated platform for user research. Our population sample consisted of 338 individuals from either the US or UK, with an approval rate of at least $99\%$. We paid the subject a total of \$4 USD for a test duration of about $25min$ in total.




\subsubsection{Hypotheses.}
Based on the analysis of visual factors and the hierarchical task analysis presented in the previous sections, we developed the following hypotheses for the study. All the hypotheses, as well as the experimental design and the pre-planned analysis, can be found in our pre-registered study \footnote{\url{https://osf.io/79ujk?view_only=63dde87519654c53abc3c06361fa05ba}}.


\begin{itemize}[noitemsep,topsep=1pt]
    \item[\textbf{H1}] Response Time: We expected alignment to have the strongest impact on speed. Therefore we expected unaligned rule representation (\texttt{SU}) to take substantially more time than feature-aligned rule representations (\texttt{SA}, \texttt{GA}, \texttt{HA}). As for the comparison between \texttt{SA}, \texttt{GA} and \texttt{HA} we expected \texttt{HA} to have an advantage over \texttt{GA} and \texttt{HA} because it enables both symbolic and graphical reading of the conditions. We expected these differences to be substantially less pronounced than the difference afforded by alignment.
    \item[\textbf{H2}] Accuracy: Our tasks are designed to focus more on efficiency than correctness. We did not expect any substantial differences in accuracy between the conditions.
    \item[\textbf{H3}] Confidence: We expected the control condition with unaligned features and symbolic encoding to have higher confidence scores due to the level of familiarity our subjects may have with this solution compared to the more unfamiliar ones we used in the other conditions.
    \item[\textbf{H4}] Subjective Effort: We expected the participants to assign to the baseline (\texttt{SU}) way higher effort scores than to the other conditions. We did not have specific expectations for \texttt{SA}, \texttt{GA} and \texttt{HA}.
\end{itemize}



\subsection{Results}
The dropout rate for those who did not finish the study was $10\%$. In the complete replies we collected, we excluded the subjects who performed the tasks too fast (<5s/question) or too slow (>100s/question for T1 or >120s/question for T2) as stated in the pre-registration plan. Then we conducted the analysis based on these samples (n=73, 78, 75, 77 for the condition \texttt{SU}, \texttt{SA}, \texttt{GA}, \texttt{HA}). 

All results are presented using absolute effect size and confidence intervals (using bootstrap $95\%$ confidence intervals) as suggested in~\cite{dragicevic2016fair}. We on purpose avoid the calculation of p-values to avoid dichotomous thinking~\cite{besanccon2019continued}. We use in their place the language of estimation driven by effect sizes and confidence intervals as suggested by Cumming in~\cite{cumming2013understanding}. For the analysis of the results, we use \texttt{SU} as a baseline, that is, the results of the other conditions will be presented in relation to the control case. An overview of the results is shown in Table~\ref{table:cohens_d}.

\subsubsection{Task 1: Prediction Estimation.}

\begin{figure*}
    \centering
    \includegraphics[width=.8\textwidth]{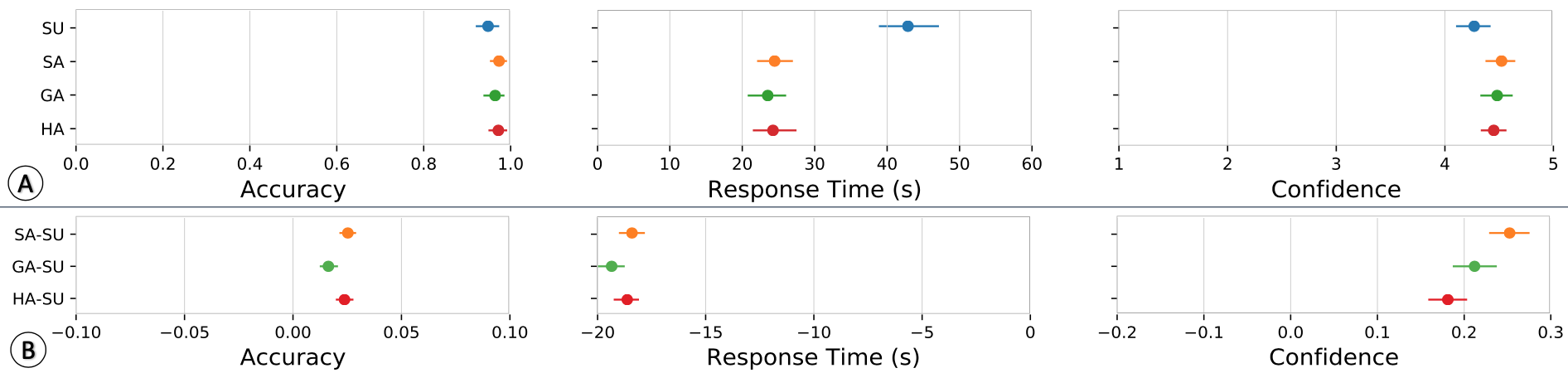}
    \vspace*{-.3cm}
    \caption{Performance of Task 1. Part (A) shows the mean values as well as $95\%$ confidence interval of accuracy, time, and confidence. Part (B) shows the absolute effect size as well as the confidence interval of the comparison between aligned conditions (\texttt{SA}, \texttt{GA}, \texttt{HA}) with the control condition (\texttt{SU}).}
    \vspace*{-.3cm}
    \label{fig:task1}
\end{figure*}

We provide the performance of Task 1 across visual conditions in Figure ~\ref{fig:task1}. Accuracy across all conditions is very similar and very close to $100\%$. And all the aligned conditions have extremely similar accuracy values with highly negligible differences. 

In terms of the efficiency, the response time with the control condition is substantially higher. The average response time for performing Task 1 with (\texttt{SU}) is over 40 seconds, while the response time with the other aligned conditions (\texttt{SA}, \texttt{GA}, \texttt{HA}) is less than 30 seconds. For this task, we observe that alignment has a stronger influence than encoding for that all the feature-aligned conditions result in similar amount of time reduction. As shown in Figure~\ref{fig:task1}B:Response Time, the participants with feature-aligned conditions (\texttt{SA}, \texttt{GA}, \texttt{HA}) spent around 19 seconds less for Task 1 on average than the participants reading rules without feature alignment (\texttt{SU}). This is a large efficiency improvement considering that users with the baseline condition spend around 40 seconds on average for Task 1. When comparing the encoding strategies over the alignment (\texttt{SA}, \texttt{GA}, \texttt{HA}) we can see that the differences are way less pronounced. Out of the 3 conditions the Graphical encoding (\texttt{GA}) seems to have a slight advantage of 0.95s faster than Symbolic encoding (\texttt{SA}) and 0.72s faster than Hybrid encoding (\texttt{HA}). \looseness=-1

Alignment also leads to an improvement in subjective confidence. Symbolic encoding brings the most improvement around $6.25\%$ (0.25/4), while graphical encoding takes the second place around $5\%$ (0.2/4), hybrid encoding around $4.5\%$ (0.18/4).  \looseness=-1

For the performance of Task 1, it is clear that a proper visual encoding together with feature alignment can substantially improve the task efficiency a lot and the accuracy can still remain at a similar level as the baseline, if not improved. We expected the participants to assign higher confidence scores to the baseline condition due to its familiar design. We actually observe that people with more efficient and more accurate rule representations also have higher subjective confidence on this prediction estimation task.

\begin{table}
    \centering
    \begin{tabular}{l| l l l | l l l| l l l | l l l }
        \hline
          \multirow{2}{*}{\textbf{Task}} & \multicolumn{3}{c|}{\textbf{Accuracy (0.00-1.00)}} & \multicolumn{3}{c|}{\textbf{Response Time (s)}} & \multicolumn{3}{c|}{\textbf{Confidence (1-5)}} & \multicolumn{3}{c}{\textbf{Subjective Effort (1-5)}}\\
          & \texttt{SA}-\texttt{SU} & \texttt{GA}-\texttt{SU} & \texttt{HA}-\texttt{SU} & \texttt{SA}-\texttt{SU} & \texttt{GA}-\texttt{SU} & \texttt{HA}-\texttt{SU} & \texttt{SA}-\texttt{SU} & \texttt{GA-SU} & \texttt{HA-SU} & \texttt{SA-SU} & \texttt{GA-SU} & \texttt{HA-SU}\\
        \hline
      \textbf{T1}  & 0.03 & 0.02 & 0.02 & -18.40 & -19.35 & -18.63 & 0.25 & 0.21 & 0.18 & \multirow{2}{*}{-0.27} & \multirow{2}{*}{-0.03} & \multirow{2}{*}{-0.10}\\
      \textbf{T2}  & 0.00 & -0.02 & 0.01 & -7.44 & -11.49 & -9.90 & 0.12 & 0.05 & -0.08 \\
        \hline
    \end{tabular}
    \caption{Absolute effect size for the comparison with the control condition (\texttt{SU}).}\vspace*{-.7cm}
    \label{table:cohens_d}
\end{table}

\subsubsection{Task 2: Prediction Characterization.}
\begin{figure}
    \centering
    \includegraphics[width=.8\textwidth]{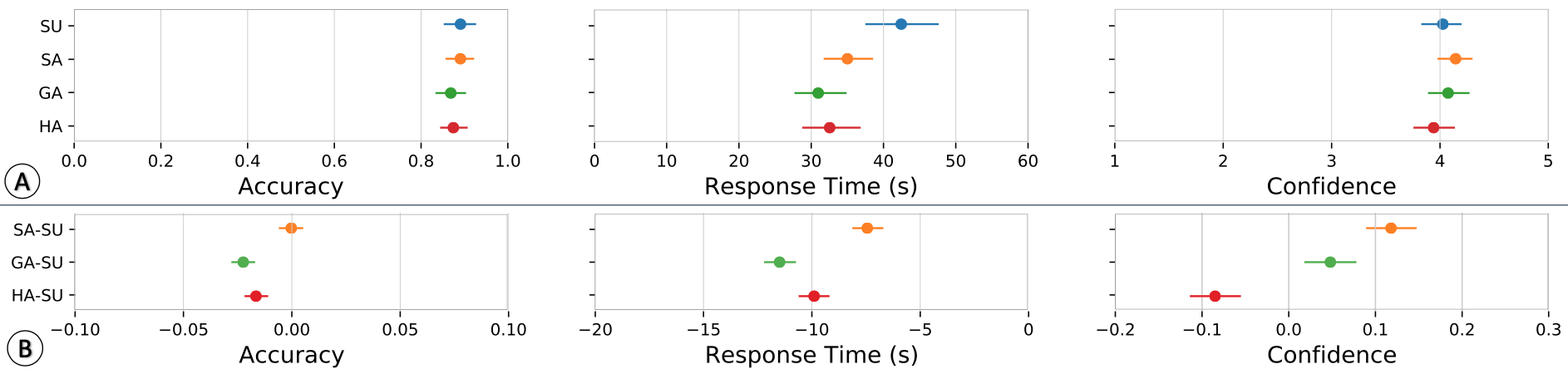}
    \vspace*{-.3cm}
    \caption{Performance of Task 2. (A) shows the mean values and $95\%$ confidence interval for each performance metric. (B) includes the absolute effect size and $95\%$ confidence interval of the comparison between  conditions \texttt{SA}, \texttt{GA}, \texttt{HA} and \texttt{SU}.} \vspace*{-.3cm}
    \label{fig:task2}
\end{figure}

In Figure ~\ref{fig:task2}, we provide, the performance of Task 2 based on three measurements. Similarly, participants with all conditions reached a relatively high accuracy. 
The influence of visual encoding and feature alignment on accuracy is more pronounced in comparison to Task 2. The three feature aligned conditions have a similar level of accuracy as the control condition.

We also observe the efficiency improvement results from feature alignment and predicate encoding. It is clear that participants who perform Task 2 with feature-aligned rules require at least 5 seconds less than with the baseline representation. Comparing the encoding strategies along with alignment, rules contain graphical encoding (\texttt{HA} and \texttt{GA}) lead to larger improvement in terms response time. More specifically, compared with the average response time from baseline (around 40 seconds), the aligned feature along with graphical encoding reduces the most, around 11.49 seconds; hybrid encoding (\texttt{HA}) results in 9.90 seconds less; symbolic encoding (\texttt{SA}) leads to around 7.44 seconds less.

As for confidence, the difference is slight. Symbolic encoding (\texttt{SA}) leads to the highest subjective confidence on average with around $3\%$ (0.12/4) higher than the confidence from the baseline. Then graphical encoding (\texttt{GA}) and hybrid encoding (\texttt{HA}) strategies lead to around $1.25\%$ (0.05/4) more and $2\%$ (0.08/4) less confidence score compared with the baseline.  \looseness=-1

For performance of Task 2, participants who use the feature aligned designs need shorter time on average while keeping a similar accuracy to the baseline representation. Condition \texttt{GA} still leads the most efficiency improvement. And participants who are using condition \texttt{SA} are more confident with their choices than those using other representations. 

In the end, subjects with the control condition (\texttt{SU)} report the highest subjective effort and subjects with other conditions report slightly less subjective effort. According to the overall subjective effort score,  \texttt{SA} leads the most improvement around $6.68\%$ (0.27/4), then \texttt{HA} $2.5\%$ (0.1/4), \texttt{GA} $0.75\%$ (0.03/4)

\section{Discussion}
Our results partially confirm some of our hypotheses. Alignment clearly has a strong impact on performance. All designs that use feature alignment have substantial improvements in response time across the two tasks, with no substantial effects on accuracy, confidence or workload. The effect on response time is very large: about a $20s$ mean difference with aligned designs on Task 1 and about $10s$ mean difference for Task 2. Such a large difference on a single inference task can have a substantial effect when considering how many of these single inferences will be carried out to make sense of a set of rules in a real-world setting.

Contrary to our expectations, the unaligned design did not lead to higher confidence. This may be due to the fact that performing the assigned tasks with the control condition requires more time than with the other conditions. If an effect of familiarity exists it is probably cancelled out by workload (time). More research is needed to understand how workload and familiarity interact. Since we did not measure familiarity directly we can only conjecture that such interaction exists. In any case, it is remarkable that the more unfamiliar visual representations we tested did not lead to substantial reduction in confidence. One open issue is whether more training and exposure to these unfamiliar designs may lead to even higher improvements than those we observed in our study.

The effect of predicate encoding is way less pronounced than the effect of alignment. Across the two tasks graphical encoding with alignment (\texttt{GA}) seems to have a slight advantage in terms of response time over the other aligned designs (\texttt{SA} and \texttt{HA}). Such an advantage is more pronounced in Task 2 where \texttt{GA} has about a mean $4s$ improvement over \texttt{SA} and about $1.5s$ over \texttt{HA}. Considering that the average duration of Task 2 for these condition is about $32s$, this is a non-negligible potential improvement of about $5-12\%$, with no substantial differences in accuracy. Regarding why such an improvement exist with \texttt{GA}, we do not have a definite explanation. Our original intuition was that the hybrid design (\texttt{HA}) would integrate the benefits of symbolic and graphical representations, but our results do not confirm this intuition. Including a symbolic representation in a predicate seems to slow down performance, even if slightly, maybe due to more time spent reading the symbols.

The condition with aligned features and symbolic (SA) encoding has a slight advantage in terms of confidence, but the effect is small and insignificant. 




Extrapolating from a single study is always difficult and for this reason we believe more studies in this space will be needed to build a more accurate understand of the effect of visual features on rule understanding. However, we feel confident we can draw some conclusions from what we have observed. The effect of alignment is extremely strong and we expect this effect to be easy to replicate and robust to many variations (e.g., different ways to encode the predicates). The effect of graphical encoding is less pronounced but not necessarily negligible. \texttt{GA} and \texttt{HA} do not seem to impact performance negatively in any conceivable way and \texttt{GA} seems to produce non-negligible performance improvement. Equipped with this knowledge we feel confident in suggesting the use of aligned layouts with graphical representations (\texttt{GA} and \texttt{HA}). They speed up performance considerably without affecting accuracy and with very negligible effects on confidence. The only downside we think designers should be aware of is that the aligned designs require more space to be presented. Since they use space for alignment they tend to be substantially less compact than the standard design where alignment is not present (see Figure~\ref{fig:test_vis} for a comparison of solution A to B, C, D).

\section{Limitations and Potential Extensions}


There are number of relevant limitations in the work we presented. First, the study covers a narrows space of conditions. As explained in Section~\ref{sec:background}, rules can vary in complexity in many ways, regardless their visual representation. What effect these factors have on rule understanding is outside the scope of our study.  Our focus has been on what we believe are the most impactful factors. Similarly, we focused on rules containing only ordinal features with a small set of values. While it is likely that our results will extend to other data types with the solutions we presented in Figure~\ref{fig:visual_factor}, more work is needed to validate this intuition. Our characterization of the design space can however be used as an aid for the exploration of novel visual representations and the development of future studies. As such, the characterization we proposed makes a substantial contribution towards future research in this area.



Second, our study focuses exclusively on rule understanding tasks but it does not explore whether the improvement in performance is reflected in an improvement in the understanding of the model described by the rules. As users make sense of models through rules, they form mental models of how the ML model works, discovering limitations and potential gaps, as well as confirming conditions in which the model works as expected. We are moderately confident that the performance improvements we have observed will be reflected in improvements in model sense making but additional experimental evidence is needed to understand the relationship between rule understanding and mental model formation. \looseness=-1

Third, our analysis of factors and the study we presented do not include interactivity. In real-world applications the visual representations we presented can be substantially enhanced by the use of interactivity. Information visualization has a long list of potentially useful interventions in this space that may make rule visualization even more powerful. Some examples of interventions include the ability to sort and filter rules and features according to different criteria as well as the ability to switch from one type of visual representation to another according to different needs. Further exploration of this space is needed to understand how interactivity may further support rule understanding and exploration. \looseness=-1

\section{Acknowledgement}
\label{sec:ackow}
We thank all the study participants and reviewers for their comments. This work was partially supported by a contract with Capital One and the NSF grant number 1928614.

\bibliographystyle{ACM-Reference-Format}
\bibliography{acmart}

\end{document}